\documentclass[aps,superscriptaddress,twocolumn,nofootinbib,12pt,tightenlines]{revtex4}

\makeatletter

\usepackage{natbib}
\usepackage{dcolumn}
\usepackage{graphicx}
\usepackage{amsmath}
\begin{document}
\title{Mechanical vs. informational components of price impact}

\author{J. Doyne Farmer}
\email{jdf@santafe.edu}
\affiliation{Santa Fe Institute, 1399 Hyde Park Road, Santa Fe, NM 87501}

\author{Neda Zamani}
\email{neda@cs.usyd.edu.au}
\affiliation{School of IT, Faculty of Science, The University of Sydney, Australia}
\affiliation{Santa Fe Institute, 1399 Hyde Park Road, Santa Fe, NM 87501}

\begin{abstract}
We study the problem of what causes prices to change.  It is well known that trading impacts prices -- orders to buy drive the price up, and orders to sell drive it down.  We introduce a means of decomposing the total impact of trading into two components, defining the {\it mechanical impact} of a trading order as the change in future prices in the absence of any future changes in decision making, and the {\it informational impact} as the remainder of the total impact once mechanical impact is removed.  This decomposition is performed using order book data from the London Stock Exchange.  The average mechanical impact of a market order decays to zero as a function of time, at an asymptotic rate that is consistent with a power law with an exponent of roughly $1.7$.  In contrast the average informational impact builds to approach a constant value.   Initially the impact is entirely mechanical, and is about half as big as the asymptotic informational impact.  The size of the informational impact is positively correlated to mechanical impact. For cases where the mechanical impact is zero for all times, we find that the informational impact is negative, i.e. buy market orders that have no mechanical impact at all generate strong negative price responses.
\end{abstract}
\maketitle

\tableofcontents

\section{{Introduction}}

What causes prices to change?  Despite a great deal of research on this subject the answer remains far from clear.  On one hand, everyone agrees that prices respond to information -- good news drives prices up and bad news drives it down.  On the other hand, prices often change even when there is little information, sometimes by large amounts \cite{French86,Cutler89}.  How much of price changes are information driven, and how much are due to other factors?

At an immediate level it is clear that trading is an important cause of price change.  When a trade is initiated by a buyer, the price tends to go up, and when it is initiated by a seller, it tends to go down.  The change in prices associated with a given trade is called price impact (or alternatively market impact), and has now been extensively studied \cite{Hasbrouck91,Hausman92,Chan93,Chan95,Farmer96,Torre97,Kempf99,Chordia01,Plerou02,Evans02,Hopman02,Potters03,Lillo03,Bouchaud04,Bouchaud04b,Lillo03c,Farmer06,Wyart06}.  Price impact says nothing about why trades are made:  They could be made because of new information or they could be made at random, for example, because a participant needs cash for reasons that are unrelated to anything going on in the market.   This is further complicated by the fact that market participants regard trading and the price changes it produces as important signals about private information \cite{Grossman80,Grossman89}.  When market participants observe a change in price, they may reason that others have private information, which may cause them to trade, setting off a cascading avalanche of price changes whose origin is difficult to ascertain.

Why do trades have price impact?  One cause is mentioned above -- they reflect information, and this information is incorporated in price changes.  There is, however, an even more fundamental reason:  When a trading order is placed it causes purely mechanical changes in prices.  The word ``mechanical" refers to the component of price changes that is deterministic, that occurs under the rules of the auction (dependent on the set of queued trading orders) even in the absence of any information.  Although the mechanical impact of trading is often discussed (see, e.g. Hopman \cite{Hopman02}), up to this point no one has offered a precise definition of what it means, or made any suggestions as to how it can be measured.

In this paper we take advantage of the fact that in most modern financial markets prices are formed using a continuous double auction (see Section~\ref{doubleAuction}).  Once the sequence of trading orders are given, the auction proceeds according to deterministic rules, acting as a deterministic dynamical system with exogenous inputs.  While the placement of trading orders may depend on complicated factors, once the sequence of trading orders is given, price formation is purely mechanical.   

We propose a definition for mechanical impact and introduce a practical method to compute it.  The mechanical impact of a trading order can be defined as the change in future prices that occurs even if no other trading orders are changed in any way.  This can be computed by introducing hypothetical alterations in the size of a trading order, and using the deterministic nature of the continuous double auction to simulate their effect on the future price sequence, holding all other aspects of the trading order sequence constant.  In contrast, the informational impact of a trading order is what is left of the total impact after the mechanical impact is removed, i.e. it is the component that depends on relationships between orders.  As discussed in Section~\ref{causality}, this can be either the causal effect that a given order has on future orders, or correlated effects between the placement of orders (e.g. due to a common cause).  More precise definitions are given in Section~\ref{decomposition}.

We find that mechanical impact behaves very differently than informational impact.  The immediate effect of placing an order is entirely mechanical.  The average mechanical impact decays to zero monotonically in time, although at a slow rate.  The long time behavior is consistent with the hypothesis of a power law, with an exponent of about $1.7$.  In contrast, informational impact grows with time and approaches a constant value, at least over the time horizon where statistics remain reliable.  On average the initial mechanical impact is about half the asymptotic informational impact.   We find that the integrated mechanical impact and the informational impact are positively correlated, except for the special case when the mechanical impact is identically zero.  In this case the informational impact is on average non-zero, with the opposite sense that one normally expects (i.e. buy orders with zero mechanical impact generate negative price impact).

This paper is organized as follows:  Section~\ref{data} gives a brief summary of the properties of our data set and gives some background information about the functioning of the continuous double auction and the London Stock Exchange.  In Section \ref{decomposition} we give more precise definitions of total price impact and its decomposition into mechanical and informational components.  We then measure the average impacts and durations for real data in Section~\ref{empiricalStudy}.  In Section~\ref{longMemory} we study the effects of long-memory in amplifying mechanical impact.  In Section~\ref{correlation} we study the correlations between mechanical and information impact. Finally in Section \ref{sec:Summary} we discuss the implications and future directions of this work.

\section{Background\label{data}}

\subsection{Continuous double auction \label{doubleAuction}}

We give a brief review of the continuous double auction, which is the most common mechanism used for trading in modern financial markets, and define some terminology that will be essential in the remainder of the paper.  ``Continuous" refers to the fact that the market is asynchronous, so that trading orders can be placed at any time, and orders are received one at a time, so there is unique ordering of events. ``Double" refers to the fact that both buyers and sellers are allowed to update their orders at will.  Orders contain both a trading quantity and a limit price, which is the worst price the trader is willing to accept\footnote{Sometimes the desired price is omitted, indicating a willingness to accept any price.  This is called a {\it market order}.  A buy market order is equivalent to a limit order with an infinite limit price, and a sell market order is equivalent to a limit price of zero.}.  The queue of unexecuted orders is called the {\it limit order book}.  A transaction is generated whenever an order crosses the prices of orders of the opposite sign, e.g. if a buy order has a higher price than the lowest priced sell order.  In this case we would say that the transaction is {\it buyer-initiated}; similarly, if a sell order crosses the best price, we say that the transaction is {\it seller-initiated}. If an order does not generate an immediate transaction, it is added to the limit order book without a transaction taking place. It is also possible to cancel an order sitting in the limit order book at any time.

Real markets have a variety of different types of possible orders that vary from market to market, but for our purposes it is possible to categorize all orders into three types of {\it events}:  {\it Effective market orders}, defined as any order or component of an order that generates a transaction; {\it Effective limit orders}, defined as any order or component of an order that does not generate an immediate transaction, and {\it effective cancellations}, defined as any removal of an order from the limit order book without a transaction taking place.  Our notion of effective events may not be in one-to-one correspondence with the actual orders that are placed.  For example, in a situation where the lowest priced sell order on the book is to sell $1000$ shares at $50$ pounds, an order to buy $3000$ shares at $50$ pounds will result in a transaction for $1000$ shares at $50$ pounds and leave a buy order sitting in the book for $2000$ shares.  This event corresponds to two effective orders, an effective market order followed by an effective limit order.

The LSE has two parallel markets, the on-book market and the off-book market.  The on-book market operates via a continuous double auction as described above, in which the limit order book is transparently visible to everyone, but the identities of those placing the orders are concealed.  The off-book market operates through a bilateral exchange in which agents contact each other via the telephone or message boards and know the identity of the agent they are trading with.  Trades in the off-book market are revealed only after they occur, and the intention to trade beforehand is only communicated to a limited circle of contacts.  For this reason the on-book market is widely regarded as the dominant force in price formation.  It accounts for more than half the number of trades and about half the trading volume.  Here we study only data from the on-book market.

\subsection{Data}

For the purpose of this study we have used the TDS (Transaction Data Service) data set for the on-book market of the London Stock Exchange (LSE).  The data set consists of records of orders placed or cancelled within a 3 year period during 2000-2002.  We study three of the most liquid stocks, Astrazeneca (AZN), Vodafone (VOD), and Lloyds (LLOY).  We will use AZN for all the figures in the paper, but we have repeated all the analyses for the other stocks as well.  Orders placed during opening and closing auctions are excluded. The resulting reduced data set contains about $570K$ transactions and $3.7M$ events for AZN,  $1.0M$ transactions and $4.0M$ events for VOD, and $600K$ transactions and $2.9M$ events for LLOY.

We have done some cleaning to reduce problems due to data errors.  Because the data set contains both a record of transactions and a record of order placements we can test to be sure that both records are consistent.  We find some problems that we do our best to correct.  E.g. there are a few cases where orders are placed that are never removed.  We remove such orders.  A more serious problem is that the sequencing of the orders is not accurate for orders that are placed within the same second.  This often results in nonsensical behavior, such as negative spreads.  When this occurs we reorder the data to eliminate these problems.  This re-ordering is not always unique.  However, this is rare, so that the overall probability that an order is out of sequence is much less than one percent.  Mis-orderings have a small effect at very short time scales (e.g. a few events), but essentially no effect on longer time scales.

In working with the data we have to deal with the problem of interday boundaries.  Because the market closes at night and over weekends, there are gaps in the data, and the behavior of the data across these gaps can be quite different than that within a trading day.  To cope with this problem we have tried two different approaches.  One is to reject any situations where we cross interday boundaries.  The other is to include situations that cross interday boundaries, but to remove price changes that occur outside the period of our analysis.  We use the latter approach for the results presented here, but we do not find that it makes a big difference in our results.

\section{Decomposing Price Impact \label{decomposition}}

\subsection{Total price impact}

Throughout this paper we work with the logarithmic midprice $p_t$, which is defined as the average of  the logarithm of the best price for a sell order and the logarithm of the best price for a buy order. The {\it total price impact} of an event at time $t$ is defined as the difference between the price just before the event and the price $\tau$ time units later, i.e., $\Delta p^T_\tau (t) = p_{t + \tau} - p_t$.  Letting $s_t$ be the sign of an event at time $t$, we can merge buy and sell orders together and define the {\it average total price impact}\footnote{The multiplication by the sign avoids problems associated with asymmetries between buying and selling.  This was introduced in reference \cite{Bouchaud04}, where it is called the response function.} as
\[
R^T(\tau) = \langle s_t (p_{t + \tau} - p_t) \rangle = \frac{1}{N} \sum_{t=1}^N s_t (p_{t + \tau} - p_t),
\]
where $N$ is the size of the sample and $\langle \rangle$ indicates a time average over time $t$. This lumps together the price impact of buy orders (positive events) and sell orders (negative events). In general these are different, but the difference is small and negligible for our purpose here.   

We have used two time units in this study:  One is \emph{event time}, where an event is an effective limit order, an effective market order, or a cancellation.  The other is \emph{transaction time}, where a transaction is defined as an effective market order\footnote{An effective market order may transact with several different orders in the limit order book.  From the point of view of transaction time we consider this a single transaction.}.  The price corresponding to time $t$ is defined to be just before the event or transaction occurs, and incremented by one immediately after that.  For AZN about 17\% of events in the data cause immediate transactions (i.e. market buy/sell events). Hence, for AZN every unit of transaction-time is the equivalent of roughly 7 event-time units.   For AZN on average one transaction time unit = 40 seconds\footnote{The definition of mechanical impact that we propose here relies on the use of natural time units, such as event time or transaction time, as opposed to real time units (seconds).  The number of events in any fixed interval of real time is highly variable, and so it is not clear how one would formulate Equation~\ref{mechanicalImpactDef} in real time.}.

\subsection{Mechanical impact }

We define the mechanical impact in terms of the change in the midprice when an event is removed, but all other events are held constant.    The idea is to measure the component of price change that is purely due to the presence of the event, excluding any effects that are caused by related changes in the sequence of events.  For the purposes of this paper we base our definition on the removal of events, but this is not the only possibility -- one can also use addition or partial additional or removal, as discussed in Section~\ref{generalizations}. 

Our definition of mechanical impact takes advantage of the fact that, under the rules of the continuous double auction, any initial limit order book and sequence of events generates a unique sequence of limit order books, which correspond to a unique sequence of midprices.  The auction $A$ can be regarded as a deterministic function
\[
b_{t+1} = A(b_t, \omega_t )
\]
that maps an event $\omega_t$ and a limit order book $b_t$ onto a new limit order book $b_{t+1}$.  The event $\omega_{t }$ can be an effective market order, limit order or cancellation, as described in Section~\ref{doubleAuction}.   For a given sequence of events $\Omega_{t+1}^{t + \tau} = \{\omega_{t+1}, \omega_{t+2}, \ldots, \omega_{t + \tau}\}$ the auction $A$ can be iterated to generate the limit order book $b_{t + \tau}$ at time $t + \tau$,
\[
b_{t + \tau} = A(b_t, \Omega_{t+1}^{t + \tau}).
\]
The continuous double auction can thus be thought of as a deterministic dynamical system with initial condition $b_t$ and exogenous input $\Omega_{t+1}$.

Each limit order book $b_t$ defines a unique logarithmic midprice $p_t = p(b_t)$.  Simplifying the notation, we write the composition of the auction operator $A$ and the midpoint price operator $p$ as $\Pi = p \circ A$.  Thus, for any initial limit order book $b_t$ and event sequence $\Omega_{t+1}^{t + \tau}$ the logarithmic midprice at time $t + \tau$ is
\begin{equation}
p_{t + \tau} = \Pi(b_t, \Omega_{t+1}^{t + \tau}).
\label{auctionDef}
\end{equation}

There are a couple of situations that deserve more discussion.  First, for the midprice to be well-defined it is necessary that the limit order book contain at least one order to buy and one order to sell.  This is not always the case.  For example for AZN, which is one of the more liquid stocks in the LSE, out of roughly 3.7 million events we observe 171 cases where the midprice is not well-defined.  Such situations become more common for less liquid stocks.  In the case where this happens, we simply say that the price is undefined.  In our empirical work we discard such cases.  

The second potentially problematic situation occurs when there is a cancellation of an order that does not exist.   The most common cause of this is the lag between the receipt of information and the time needed to react to implement a cancellation, which can result in an order being executed before it can be cancelled.
There are 231 cases of this for AZN during the period of our sample.  When this occurs the exchange simply ignores the cancellation.  We do the same, treating the cancellation as a null event, i.e. one that leaves the limit order book unchanged. 

For the purposes of this paper we define the mechanical impact of an event $\omega_t$ in terms of its effect on the price under its hypothetical removal.  More precisely, making use of Equation~\ref{auctionDef},  we define the {\it mechanical impact} $\Delta p^M_\tau (t)$ as
\begin{equation}
\Delta p^M_\tau (t) = \Pi(b_t, \Omega_{t+1}^{t + \tau}) - \Pi(b_{t-1}, \Omega_{t+1}^{t + \tau}).
\label{mechanicalImpactDef}
\end{equation}
In other words, the mechanical impact of event $\omega_t$ at time $t + \tau$ is the difference between the real price $\Pi(b_t, \Omega_{t+1}^{t + \tau})$ and the hypothetical price $\Pi(b_{t-1}, \Omega_{t+1}^{t + \tau})$ when $\omega_t$ is removed,  but the subsequent sequence of events is left the same.  The real price contains both the informational and mechanical impact, while the hypothetical price contains only the informational impact, so that under subtraction only the mechanical impact remains.  This isolates the part of the price impact that is ``purely mechanical", in the sense that it is generated solely by the effect of placing an order in the book and observing its effect under the deterministic operation of the continuous double auction.  Although we have not been able to prove this, we conjecture that when $\omega_t$ is a buy order ($s_t = +1$) the continuous double auction guarantees that $\Delta p^M_\tau \ge 0$ for all $\tau$, and for sell orders $(s_t = -1$) $\Delta p^M_\tau \le 0$.  In our empirical investigations we have not seen any exceptions to this conjecture.  For $\tau \le 0$ the mechanical impact is by definition zero.

Removing an order to compute the mechanical impact increases the number of cancellations of orders that do not exist.   For example, suppose that in the real event sequence the buy market order $\omega_t$  removes a sell limit order $\hat{\omega}$.  When we generate the hypothetical sequence by removing $\omega_t$, $\hat{\omega}$ is left in the book, so that the best ask price remains lower than it did in the real sequence.  This can cause a subsequent buy limit order to be executed that would otherwise have remained in the book and later been cancelled.  When this occurs, as already stated, we treat this as a null event.  For AZN, for example, we find that about $35\%$ of the time removing $\omega_t$ generates at least one cancellation of a nonexistent order, i.e. $19\%$ of the time there is one such event, $7\%$ of the time two such events, etc., for an average of $0.73$ such events per removal.

In the hypothetical series we also observe a slight increase in the number of undefined prices.  For AZN, for example, this happens in $0.01\%$ of the cases, in contrast to $0.003\%$ for the real series.  We omit these cases from our analysis.

Finally, we need to deal with the possibility that the assignment of effective orders might alter the number of events in the hypothetical series.  This can happen, for example, because an order that is fully executed in the real series, and was therefore an effective market order, is now only partially executed, and so becomes an effective market order followed by an effective limit order.  We deal with this by preserving the alignment of the hypothetical and real data series based on the real events.  This prevents the possibility of a persistent misalignment which could create a persistent artificial price difference between the two series.  In any case we find that such situations are rare.

\begin{figure}[ptb]
\includegraphics[scale=0.55]{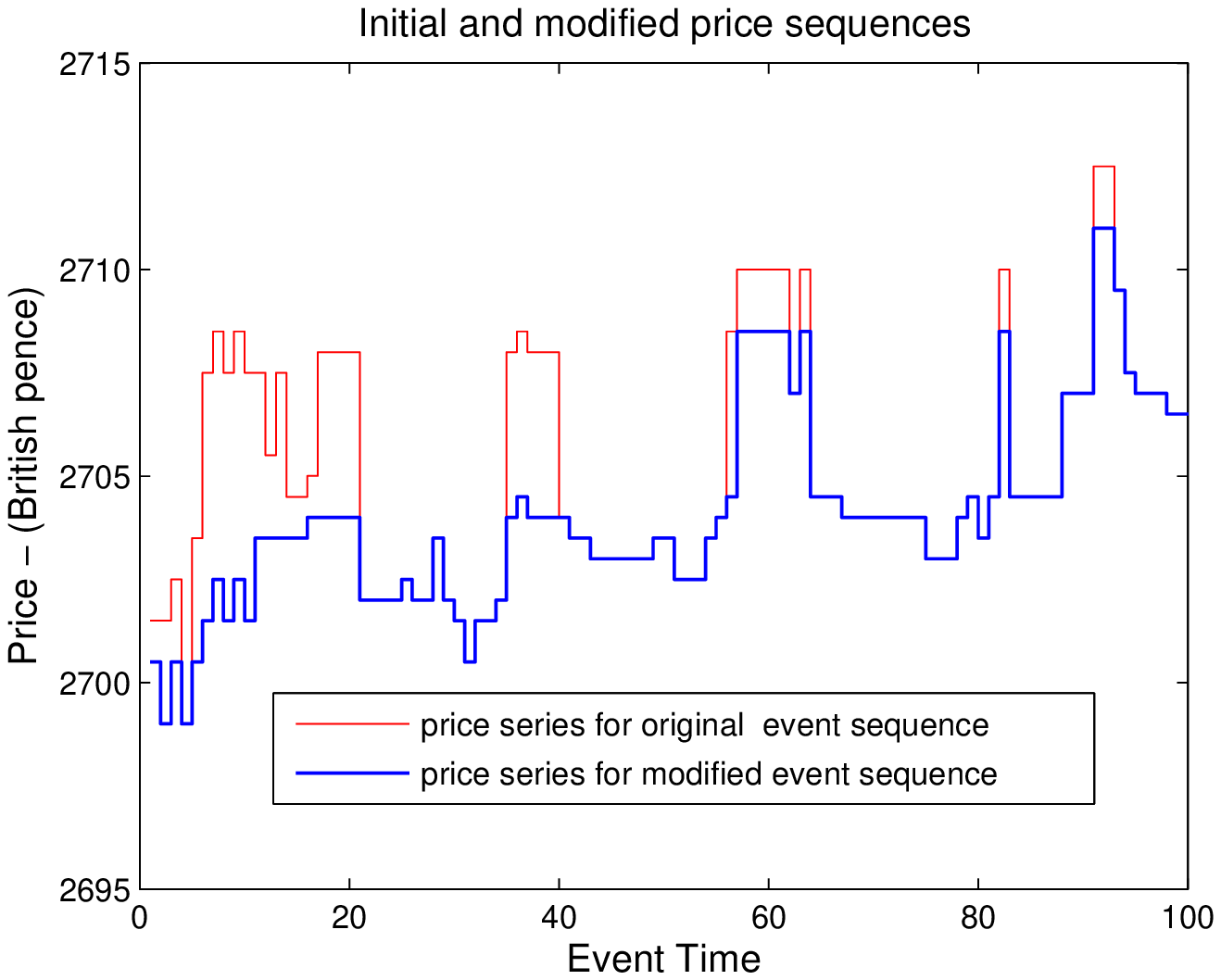}
\includegraphics[scale=0.55]{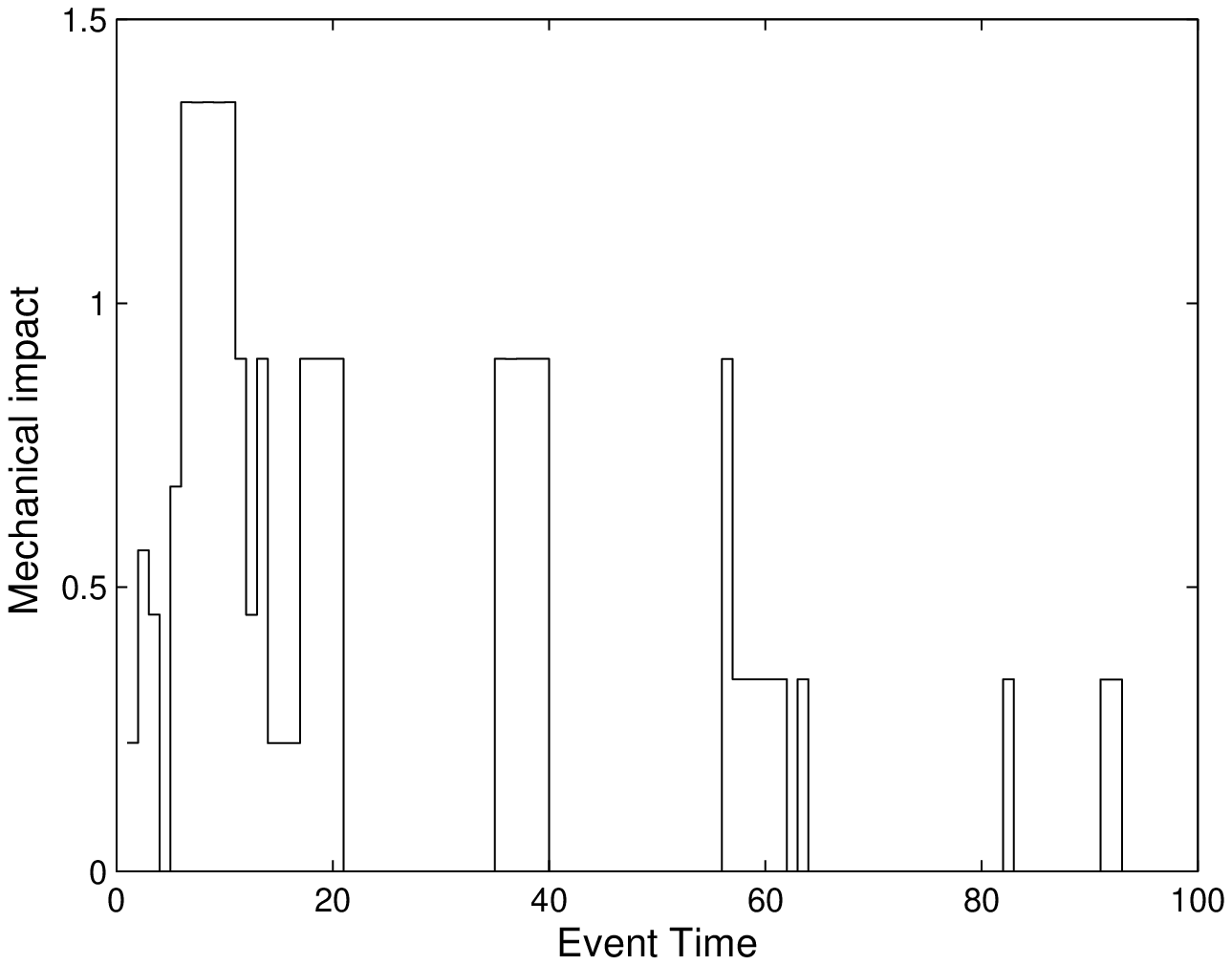}
\caption{(a) Initial and modified price sequences for the removal of a buy market order and
(b) the resulting mechanical impact (which is just the difference between the two price series in (a)). Note that this figure and all other figures in this paper are based on the stock AZN.  Prices are in pence and impacts are in units of average spread. \label{cap:DI-zoomed-example}}
\end{figure}

\begin{figure}[ptb]
\vspace{-.4in}
\includegraphics[scale=0.52]{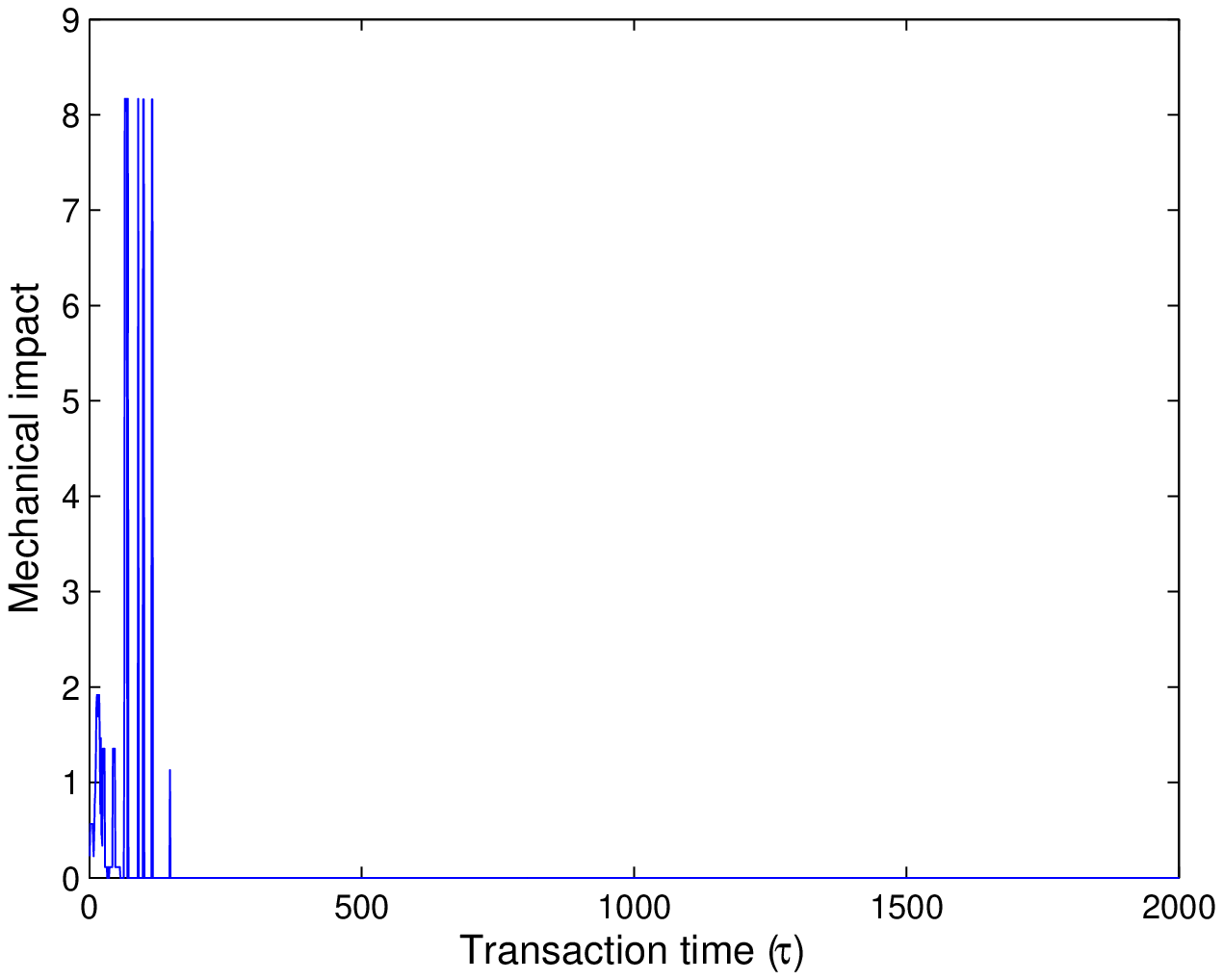}
\vspace{-.2in}
\includegraphics[scale=0.52]{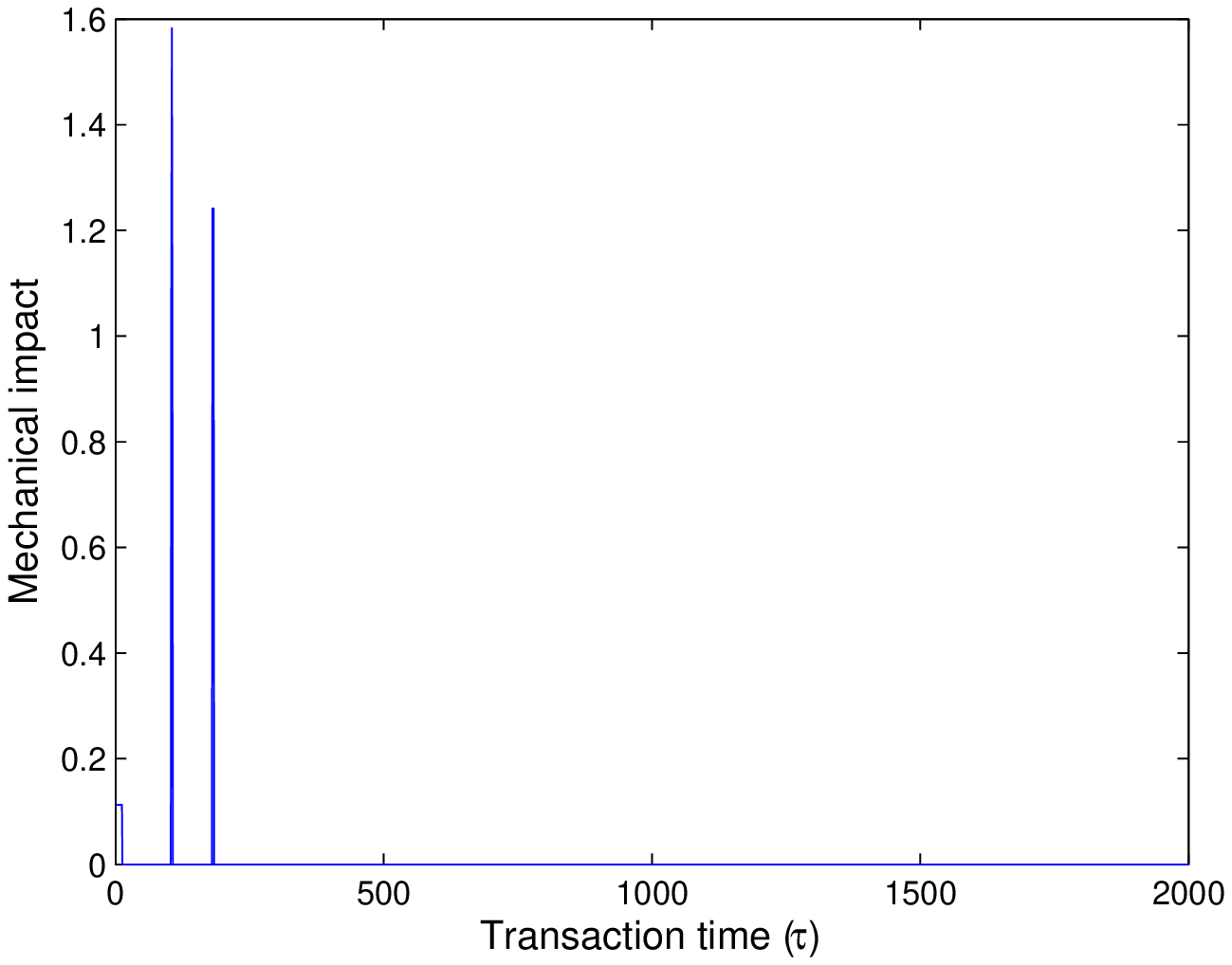}
\vspace{-.2in}
\includegraphics[scale=0.52]{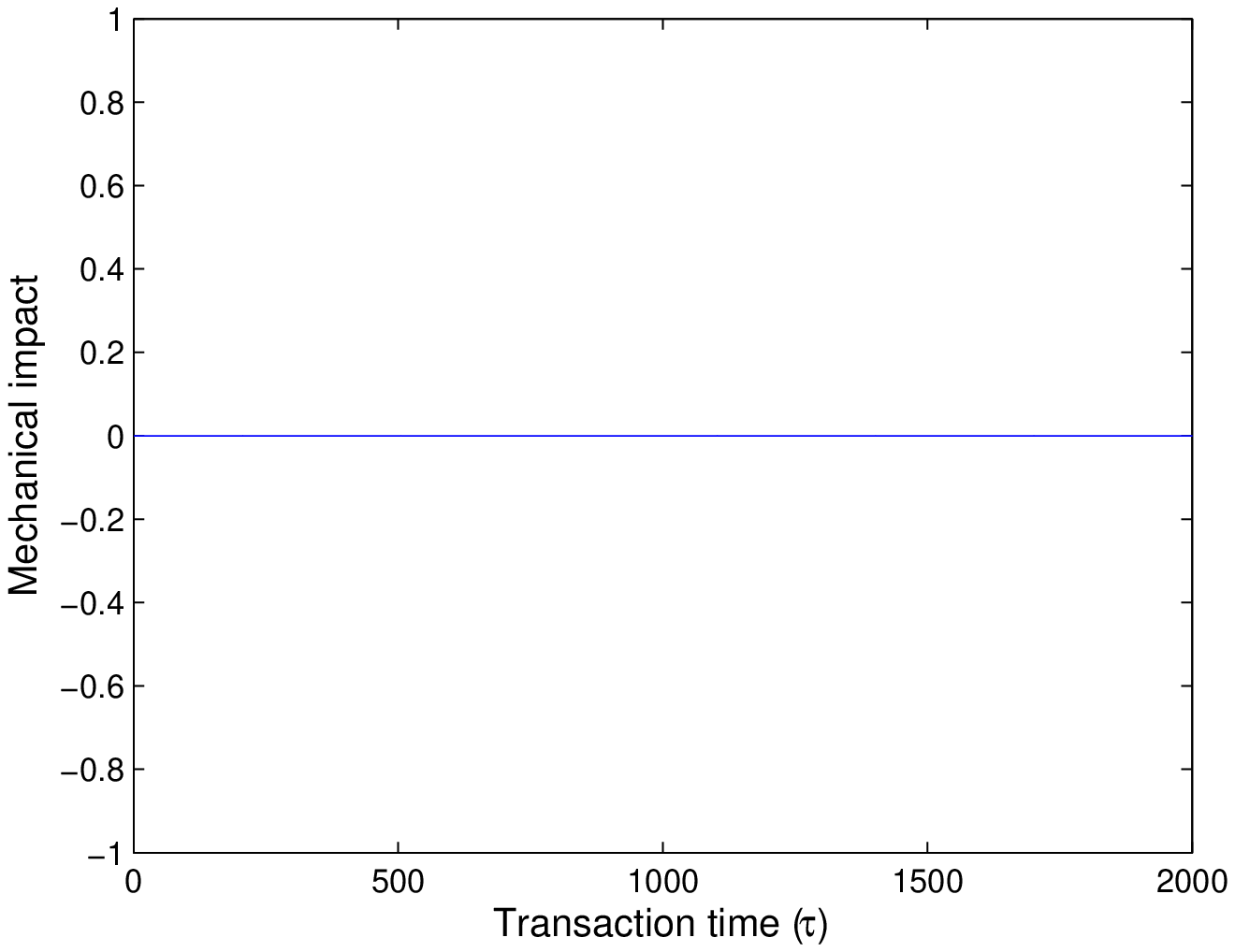}
\vspace{-.2in}
\includegraphics[scale=0.52]{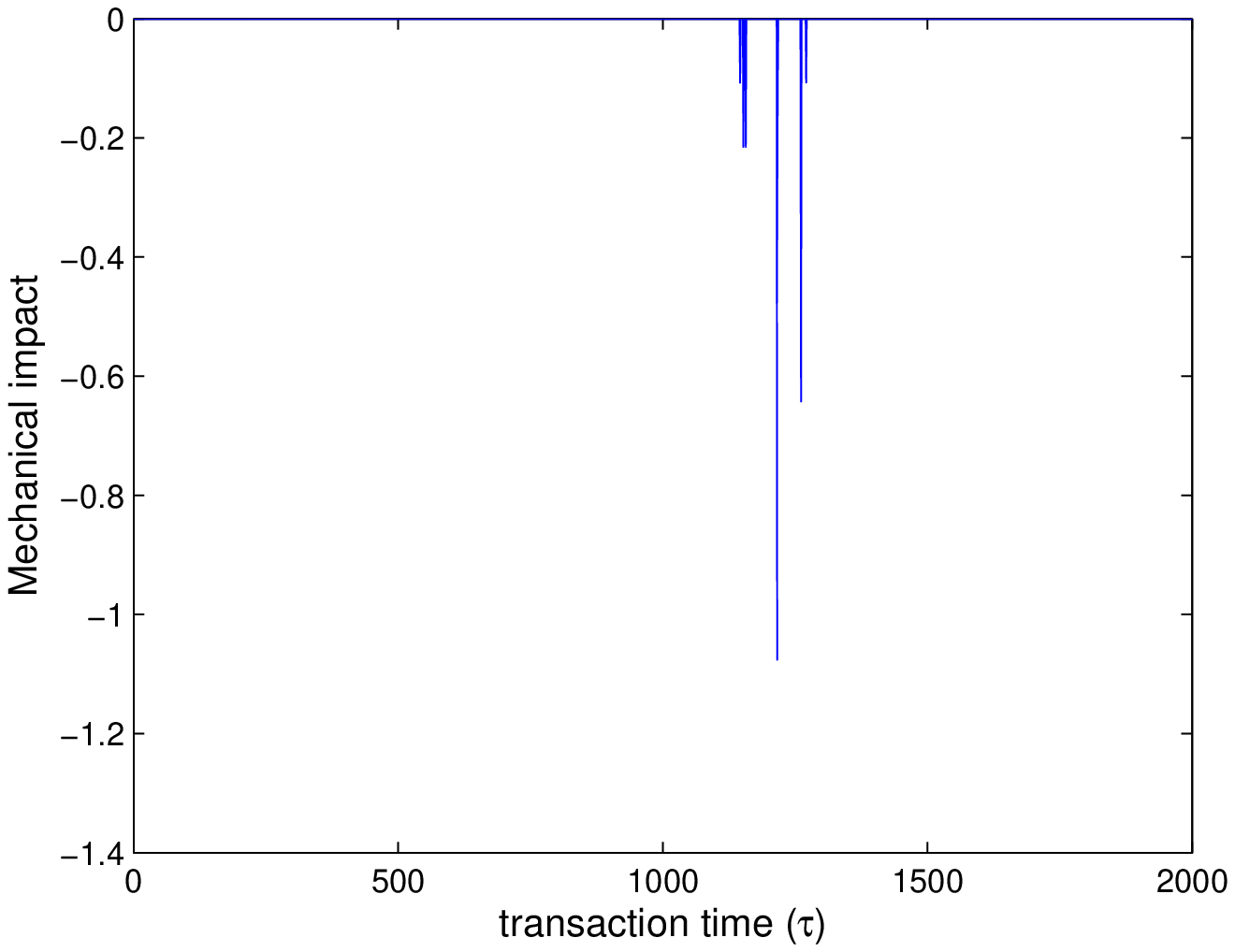}

\caption{Examples of mechanical price impact (in units of average spread) in transaction time.  The top two plots correspond to buy orders and the bottom two to sell orders.\label{cap:DI-examples}}
\end{figure}

In this paper we study only the impact of effective market orders.  In Figure \ref{cap:DI-zoomed-example} we show a typical example comparing a real price sequence to a hypothetical price sequence with a buy order removed\footnote{The TDS data set comes with a series of bids and offers computed by the exchange.  These do not always match the prices that we compute by applying the continuous double auction algorithm to the event series.  For consistency of comparison between the real and hypothetical series, we use the latter.}.
It is clear that the mechanical impact is highly variable. To give the reader a feeling for the variety of possibilities, in Figure \ref{cap:DI-examples} we show a set of four examples of the mechanical impact of effective market orders\footnote{In this and the remaining figures we compute impacts in units of the average spread.  For each stock and the full time period of each data set, the average spread is computed by sampling just before transactions.} .
The mechanical impact is highly variable.  In some cases there is an initial burst of mechanical impact, which dies to zero and then remains there. In some cases there are long gaps in which the impact remains at zero and then takes on nonzero values after more than a thousand transactions. In other cases there is no mechanical impact at all.  

\subsection{Informational impact}

The {\it informational impact} is defined as the portion of total impact that cannot be explained by mechanical impact, i.e. 
\[
\Delta p^I_\tau = \Delta p^T_\tau - \Delta p^M_\tau.
\]
Whatever components of the total impact not explained by mechanical impact must be due to correlations between the order $\omega_t$ and other events.  With the data we have it is impossible to say whether the placement of the order $\omega_t$ causes changes in future events $\Omega_{t+1}$, or whether the properties of $\Omega_{t+1}$ are correlated with those of $\omega_t$ due to a common cause.  In either case, changes in price that are not caused mechanically must be due to information -- either the information contained in $\omega_t$ affecting $\Omega_{t+1}$, or external information affecting both $\omega_t$ and $\Omega_{t+1}$.  See the discussion in Section~\ref{causality}.

\section{Empirical study of impact\label{empiricalStudy}}

In this section we perform statistical analysis of average properties of price impact. For the remainder of this paper we will only study the impacts of effective market orders, deferring the problem of studying effective limit orders and cancellations.

\subsection{Average impact}

\begin{figure}[ptb]
\includegraphics[scale=0.55]{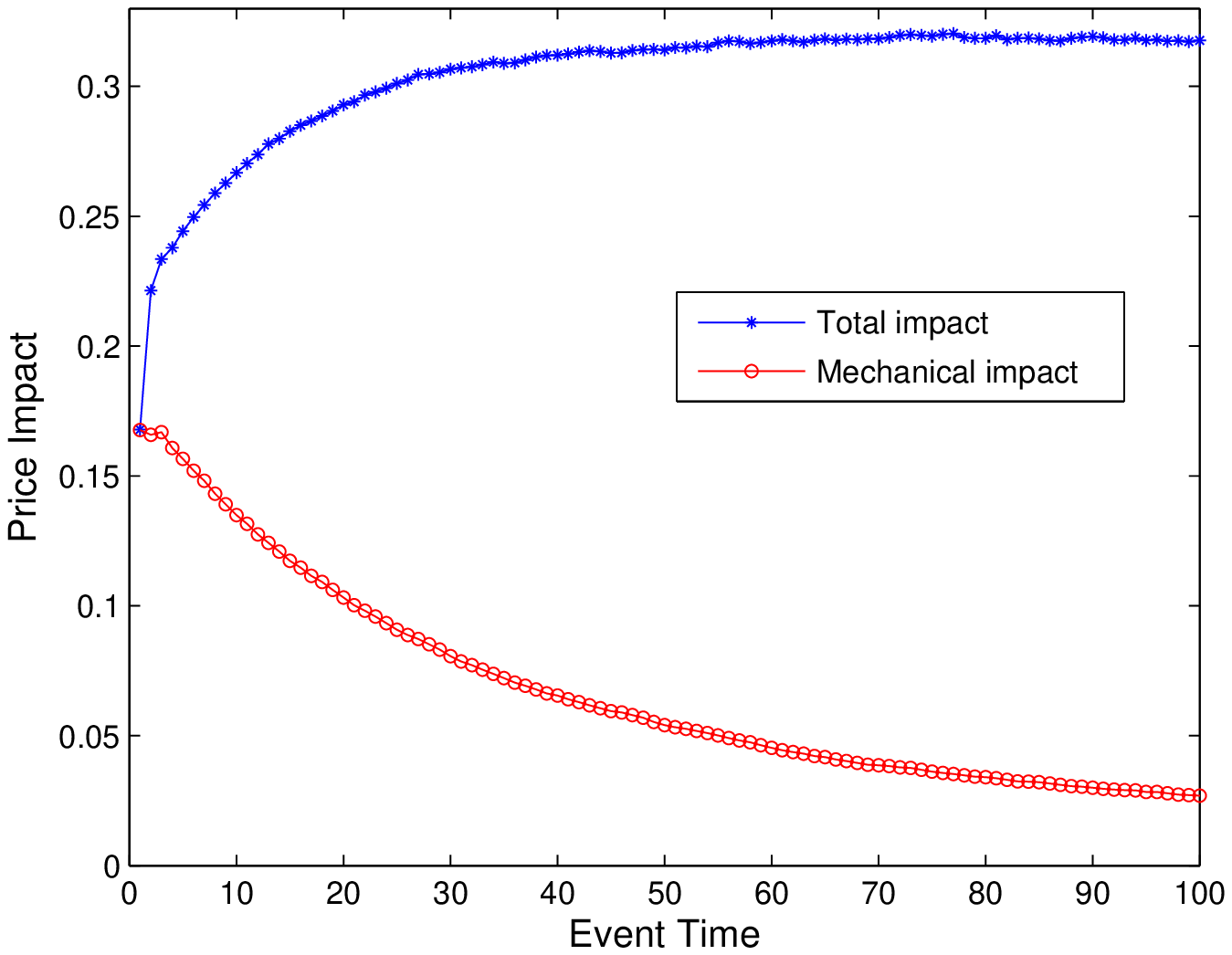}
\includegraphics[scale=0.57]{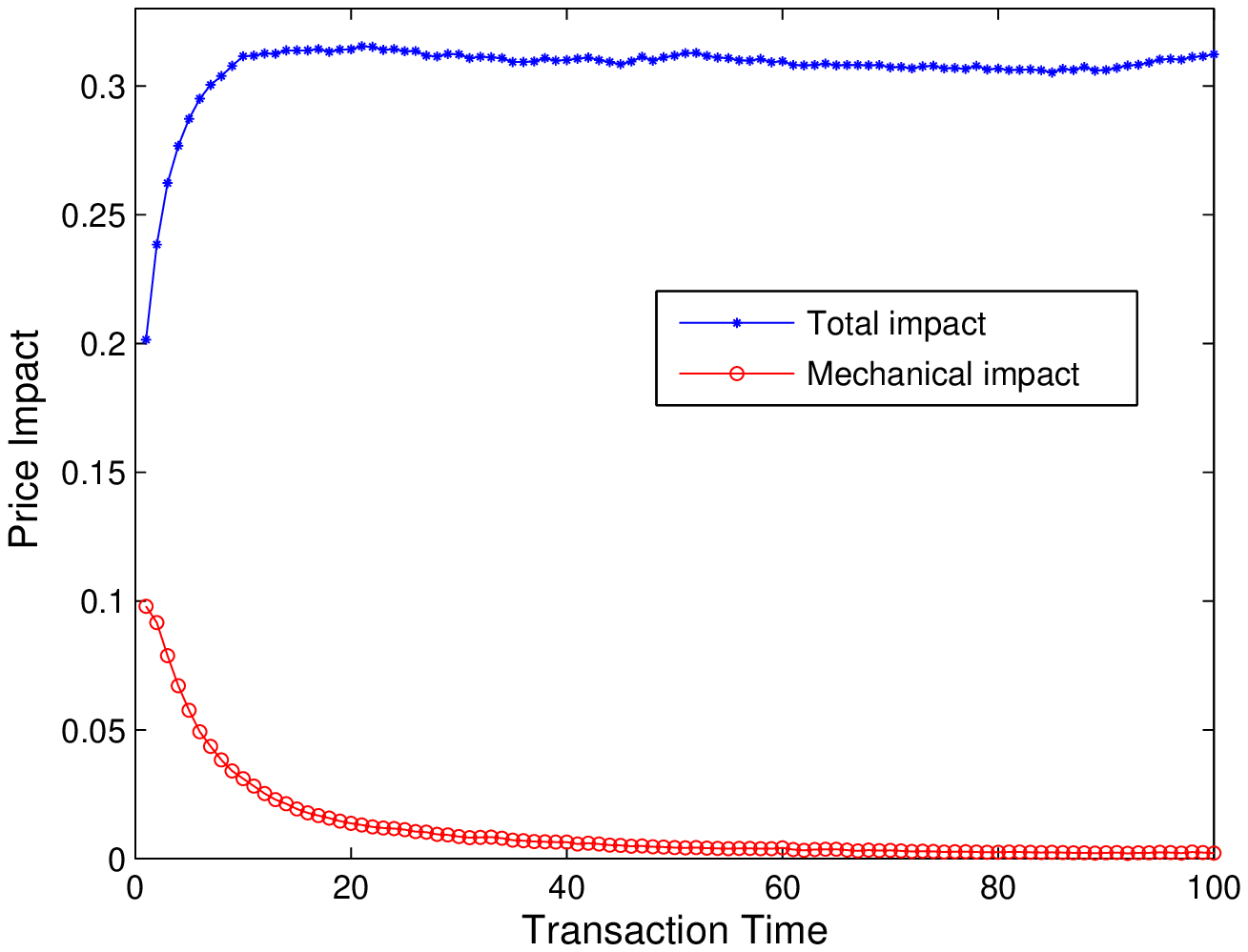}
\caption{Average mechanical impact $R^M(\tau) = \langle s_t \Delta p^M_\tau (t) \rangle_t$ (red squares) and total impact $R^T(\tau) = \langle s_t \Delta p^T_\tau (t) \rangle_t$ (blue stars), in units of the average spread, plotted in (a) event time and (b) transaction time.  \label{avImpact}}
\end{figure}

The first property we study is the average impact as a function of time. In Figure~\ref{avImpact} we compare the average mechanical impact $R^M(\tau) = \langle s_t \Delta p^M_\tau (t) \rangle_t$ and the average total impact $R^T(\tau) = \langle s_t \Delta p^T_\tau (t) \rangle_t$.
For event time the total impact and mechanical impact are by definition the same at $\tau = 1$.  This is because in moving from $\tau = 0$ to $\tau = 1$ the only event that affects the price is the reference event $\omega_t$ --  alterations in $\Omega_{t+1}$ cannot effect $\Delta p^T_1$.  For larger values of $\tau$ the mechanical impact decreases and the informational impact increases.  Over the timescale shown here (100 events), when measured in units of the average spread, the mechanical impact is initially about $0.17$, and then decays monotonically toward zero. In contrast the total impact increases toward what appears to be an asymptotically constant value slightly greater than $0.3$.  This the source of our statement that the initial value of mechanical impact is about half the asymptotic value of the total impact.  Similar results are observed for VOD and LLOY.

Figure~\ref{avImpact}(b) shows the same behavior in transaction time. In this case the mechanical impact and total impact diverge immediately at $\tau = 1$.  This is not surprising, since one increment of transaction time is equivalent to roughly $6.8$ increments of event time.  The initial gap between the total and the mechanical impacts in transaction time is roughly, but not exactly consistent with the results in event time\footnote{It is not possible to superimpose the event time and transaction time curves by simply rescaling the time axis.  The reason is somewhat subtle.  It depends on the fact that transactions on average have bigger effects on prices than other events. Because we create a hypothetical sequence by removing market orders, each event time interval always begins with a transaction, but between $6.8$ events there can be more than one transaction.  For AZN on average there are $1.2$ transactions. As a result, if one simply rescales the time axis by a factor of $6.8$, the average mechanical impact in event time is greater than that in transaction time.}  at $\tau = 6.8$. Once again, except for statistical fluctuations the total impact appears to approach a constant value out to $\tau = 100$.

The total impact of market orders has received extensive study in several different markets \cite{Hasbrouck91,Hausman92,Farmer96,Torre97,Kempf99,Chordia01,Plerou02,Evans02,Hopman02,Potters03,Lillo03}.  Here we focus on the behavior as a function of time \cite{Chan93,Chan95,Bouchaud04,Bouchaud04b,Lillo03c,Farmer06,Wyart06}.  Our results on total impact are consistent with previous results.  We find that total impact builds with time and appears to approach a constant, up to the point where statistical fluctuations make the results questionable.

\begin{figure}[ptb]
\includegraphics[scale=0.6]{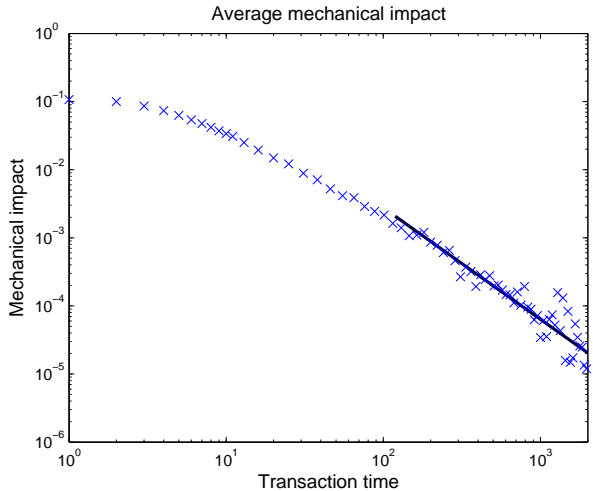}
\caption{Average mechanical impact in units of the average spread, plotted in double logarithmic scale.  The x's are the data and the solid line shows a power law fit to the tail.
\label{cap:DI-Power-law-fit}}
\end{figure}

In sharp contrast to the total impact, the mechanical impact decays toward zero. To get a better view in Figure~\ref{cap:DI-Power-law-fit} we plot the average mechanical impact for times up to 2000 transactions in log-log scale\footnote{To make the fluctuations in the data clearer in this and other figures, we have placed x's at diminishing intervals of time to get a better view in log-log scale.}. 
We fit a power law of the form $K\tau^{-\lambda}$ to what we subjectively deem to be the asymptotic region of the tail.  For AZN we find $\lambda = 1.6$, for VOD $\lambda = 1.8$, and for LLOY $\lambda = 1.7$.   Given that the scaling region is only over a little more than an order of magnitude, this is certainly not convincing evidence that the average mechanical impact scales as a power law.  Nonetheless, plotting in semi-log coordinates makes it quite clear that the decay is slower than exponential. We have not attempted to put any error bars on these estimates because they are difficult to assign\footnote{In addition to statistical fluctuations, there are problems caused by the slow convergence to asymptotic scaling. Careful testing of the power law hypothesis and proper assignment of error bars for scaling exponents is beyond the scope of this paper.}. These should just be viewed as representative values.

It is clear from simple theoretical arguments that the mechanical impact must decay to zero. From the definition of Equation~\ref{mechanicalImpactDef} the only difference between the real price series and the hypothetical series with the order removed is the initial order book $b_t$. As orders disappear from the order book the difference between the real and the hypothetical price series decreases.  Once there are no orders that were in the original order book there is no longer any memory of the initial condition, and there will no longer be any difference between the real and the hypothetical price series.  Previous studies suggest that the distribution of times for an order to remain in the book before being removed by a transaction decays as a power law with an exponent near $1.5$, and that the distribution of times to cancellation decays as a power law with an exponent near $2$ \cite{Challet02,Mike05}.  These both suggest that we should expect an asymptotic power law decay of the average mechanical impact, though the precise argument linking these is not clear.

Under the rules of the LSE no order can persist for more than one month, which would seem to imply an upper bound on the persistence of the mechanical impact.  However, it is quite common for orders to be immediately replaced, so that from an effective point of view some orders can persist for very long times, e.g. six months or more \cite{Zovko02}. For practical reasons related to limitations in computation time we have not measured the mechanical impact for time intervals longer than 2000 transactions, corresponding to a period of about two trading days.  This doesn't appear to be an important restriction, since the average mechanical impact at $\tau = 2000$ is less than three orders of magnitude less than its initial value, and events where the mechanical impact is non-zero for $\tau > 2000$ are fairly rare.

As already mentioned, the hypothetical sequences contain more cancellations of nonexistent orders, which are treated as null events.  We find that on average removals that generate more null events have larger mechanical impact.  This is not surprising, since the generation of null events implies larger perturbations in the limit order book.  (See the example discussed in the paragraph following Equation~\ref{mechanicalImpactDef}). 

\subsection{Duration and size}

\begin{figure}[ptb]
\includegraphics[scale=0.6]{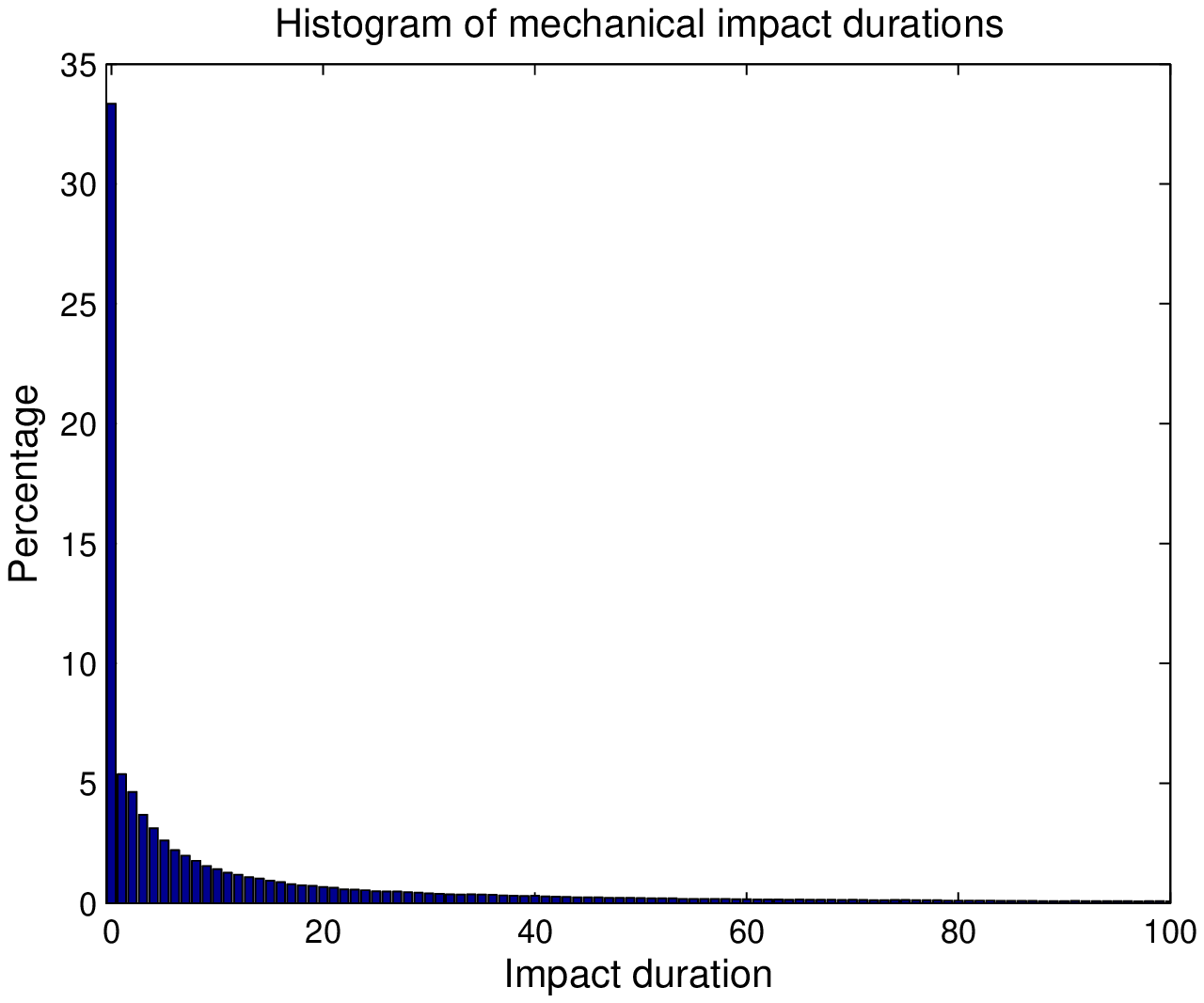}
\includegraphics[scale=0.58]{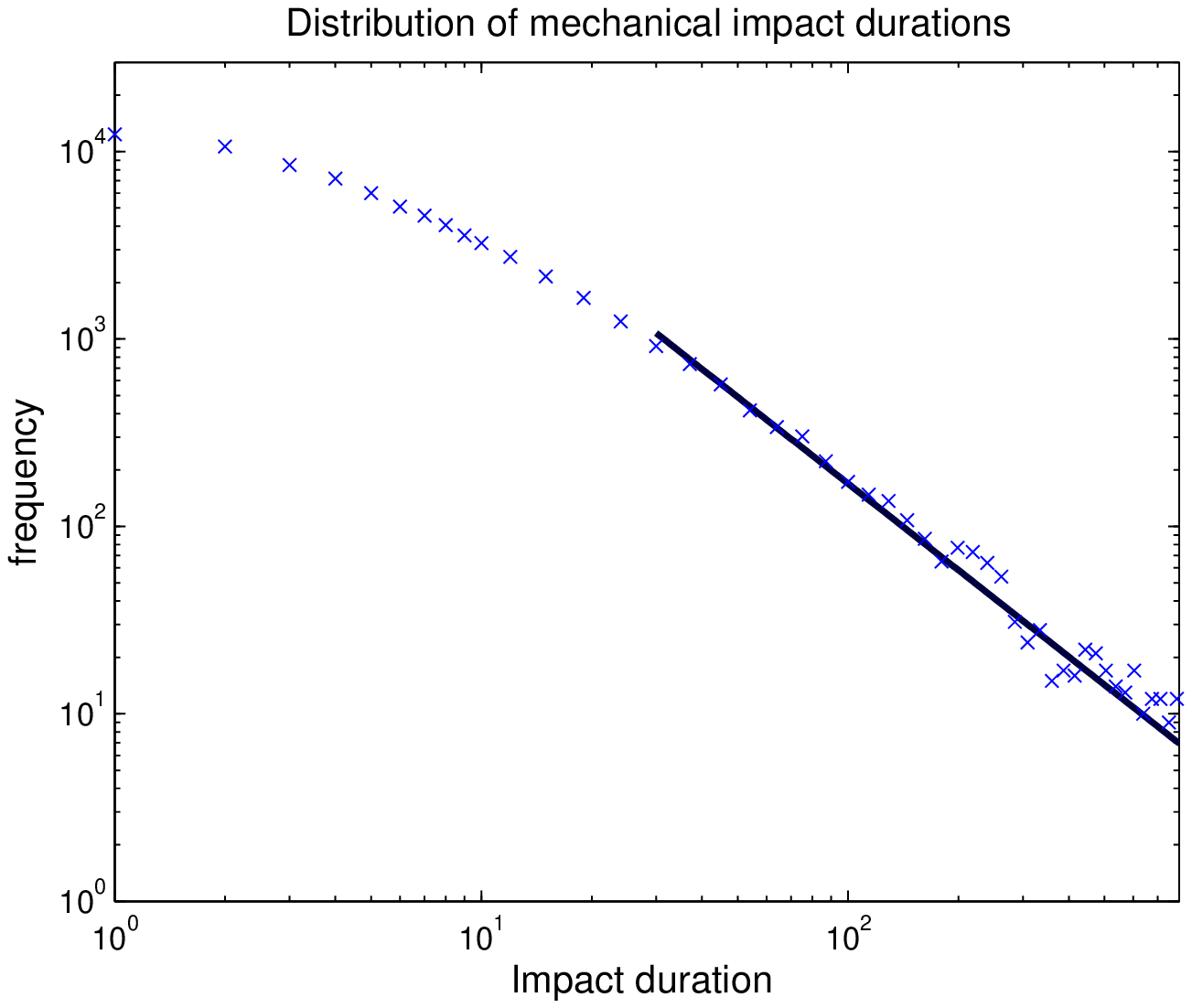}
\caption{Distribution of mechanical impact durations.  The top panel is the probability of each duration, and the bottom panel covers a longer time period on double logarithmic scale.  The x's correspond to the empirical measurements, and the line to a fit using ordinary least squares. \label{cap:DI-dur-hist}}
\end{figure}

We define the {\it duration} of the mechanical impact of an order as the largest $\tau$ for which that order has a non-zero impact.  Figure \ref{cap:DI-dur-hist} shows a histogram of the durations in transaction time. 
The most common duration is zero -- in transaction time about $33\%$ of the events have no impact at all .  The duration probability decreases rapidly and roughly monotonically.  To get a better view in Figure~\ref{cap:DI-dur-hist}(b) we show this for a longer time period in double logarithmic scale and fit a power law $K \tau^{-\delta}$ over what we subjectively deem to be the tail.  For AZN we estimate $\delta = 1.5$, for VOD $\delta = 1.7$ and for LLOY $\delta = 1.7$.

\begin{figure}[ptb]
\includegraphics[scale=0.6]{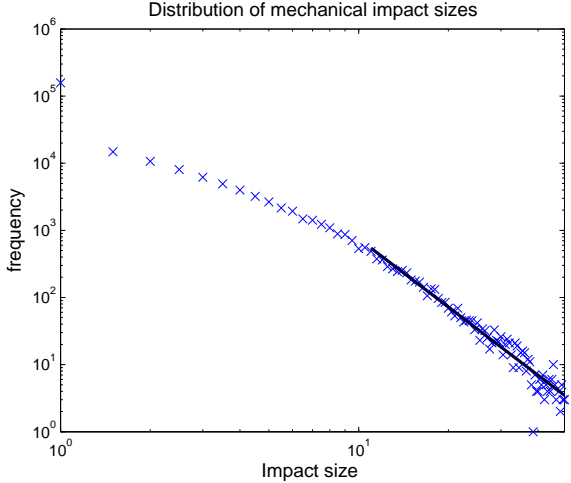}
\caption{The distribution of integrated mechanical impact sizes $S^M_\tau (t)$, with $\tau = 2000$.  The x's are the real data and the solid line is a fit to a power law.  The units of the $x$ axis are price $\times$ time, where price differences are measured in units of average spread and time is measured in transactions.  \label{cap:DI-size-PlawFit}}
\end{figure}

Because mechanical impact is transitory, we define a notion of size in terms of the impact integrated over time.   The integrated size $S^M_\tau (t)$ up to time $\tau$ associated with the impact event at time $t$ is defined as
\[
S^M_\tau (t) = \sum_{i=1}^\tau s_t \Delta p^M_i (t).
\]
The distribution of integrated sizes is shown in Figure~\ref{cap:DI-size-PlawFit} in double logarithmic scale.   
If we fit a power law of the form $K\tau^{-\alpha}$ for AZN we find $\alpha = 3.3$, for VOD $\alpha = 3.4$, and for LLOY $\alpha = 3.4$.  Given the small size of the scaling region and the large size of the scaling exponents, it is not at all clear that the integrated impact size asymptotically scales as a power law. 

\section{Amplification by long-memory\label{longMemory}}

The effect of mechanical impact on prices is amplified by the long-memory of the order signs $s_t$.  As we explain in more detail below, the long-memory of order signs refers to the strong tendency of buy orders to be followed by more buy orders, and sell orders to be followed by more sell orders.  When we compute mechanical impact we subtract the price sequence associated with two series of orders, both of which contain long-memory.  The long-memory is thus removed.  Nonetheless, since buy orders generate positive mechanical impacts and sell orders generate negative mechanical impacts, the long-memory of orders amplifies the mechanical impact of any given order.  In this section we make all this more precise, compute the amplification due to long-memory, and show that it is not strong enough to create a persistent impact from the transitory nature of individual mechanical impacts.

\begin{figure}[ptb]
\includegraphics[scale=0.6]{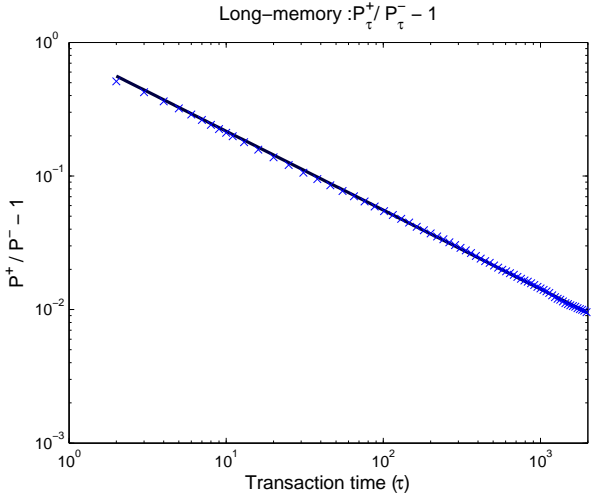}
\caption{An illustration of the long-memory of order signs.  The ratio $P^+_\tau / P^-_\tau - 1$ is plotted on double logarithmic scale (x's) and compared to a power law (line) with exponent $\gamma = 0.59$. \label{cap:P-pos-neg}}
\end{figure}

To make the definition of the long-memory of order signs more precise, define 
\begin{eqnarray}
\nonumber
P^+_\tau) = P(s_t = 0 ~\& ~ s_{t + \tau} = 0)\\
\nonumber
 + P(s_t = 1 ~\& ~ s_{t + \tau} = 1)
\end{eqnarray}
as the probability that transactions at time $t$ and $t + \tau$ have the same sign and similarly $P^-_\tau$ as the probability that they have the opposite sign.  For all the stock markets examined so far (Paris, London, NYSE) the sequence of signs has long-memory, i.e. the autocorrelation function decays as a power law $\tau^{-\gamma}$ \cite{Bouchaud04,Lillo03c,Bouchaud04b,Farmer06}.  This also implies that $P^+_\tau / P^-_\tau  - 1 \sim \tau^{-\gamma}$.  The exponent $\gamma$ appears to vary somewhat depending on the market and the stock, but it is consistently less than one.  Long-memory is illustrated for AZN in Figure~\ref{cap:P-pos-neg}.
Lillo, Mike, and Farmer \cite{Lillo05b} hypothesized that long-memory is caused by strategic order splitting and presented results supporting the hypothesis.  More recent results that make use of transaction identity codes strongly support this \cite{Vaglica06}.

The long-memory of order signs creates a puzzle because it naively suggests that prices should be inefficient.  Figure~\ref{cap:P-pos-neg} makes it clear that order signs are predictable based on their past history.  Given that buyer-initiated trades have positive price impact and seller-initiated trades have negative impact, this suggests that price movements should also be predictable.  This is not the case:  price changes are essentially uncorrelated.  

Two explanations have been offered to explain how transactions can have long-memory while prices are efficient.  One is due to Bouchaud et al. \cite{Bouchaud04,Bouchaud04b}.  They postulate the existence of a bare impact function $G(\tau)$, such that the total impact is the sum of the bare impact of each trade.   If $G(\tau)$ decays as a power law $G(\tau) \sim \tau^{-\beta}$, providing $\beta = (1 - \gamma)/2$, due to the long-memory the bare impacts will accumulate so that the total impact asymptotically approaches a constant, i.e. the decay of the bare impact and the amplification due to long-memory cancel each other.  Providing the constant is less than half the spread (which is observed in practice), the market is efficient \cite{Wyart06}.  

The temporary nature of the average mechanical impact response function shown in Figure~\ref{avImpact} suggests that it might provide a fundamental explanation for the decay of the bare impact $G(\tau)$.   However, the average mechanical impact decays too fast for this to be true.  According to the formula of the previous paragraph, using $\gamma = 0.6$ implies a bare propagator exponent of $\beta = 0.2$.   This is much smaller than the exponent $\lambda = 1.6$ measured for mechanical impact.  Thus it seems that this cannot be the explanation.  Instead, the full explanation involves the existence of a liquidity imbalance between buying and selling\footnote{The hypothesis of a liquidity imbalance between buying and selling was offered by Lillo and Farmer\cite{Lillo03c} and modified and demonstrated to be effective by Farmer et al. \cite{Farmer06}.  When buyer-initiated transactions become more likely, the liquidity for buying increases and the liquidity for selling decreases (though with some time lag), making the price responses to buy orders smaller on average than those to sell orders.  This damps the effect of the long-memory and keeps the market efficient.}.  Nonetheless, the decaying nature of the mechanical impact may play an important role in making the bare impact temporary.  The details of this remain to be worked out.

\begin{figure}[ptb]
\includegraphics[scale=0.6]{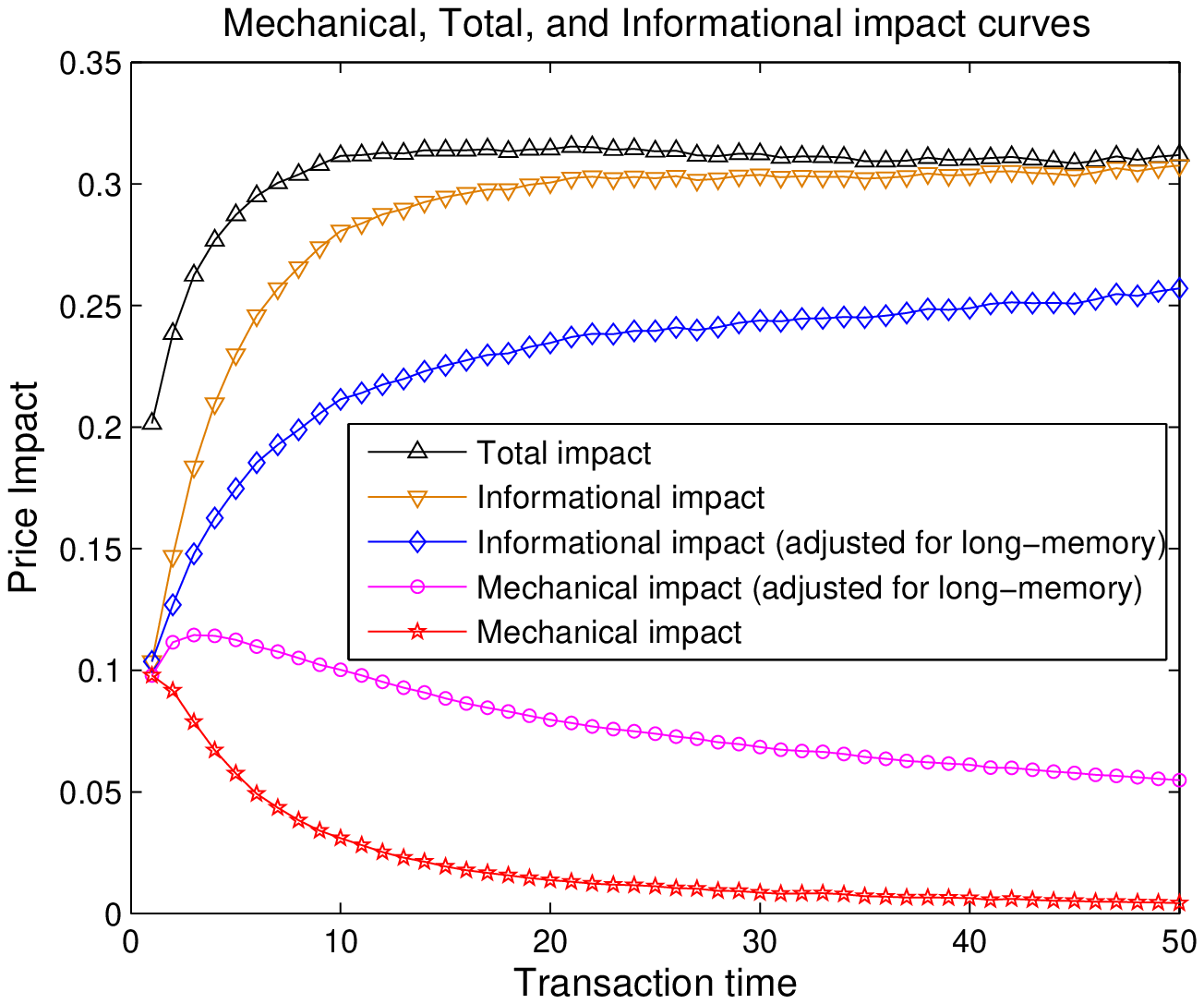}
\includegraphics[scale=0.6]{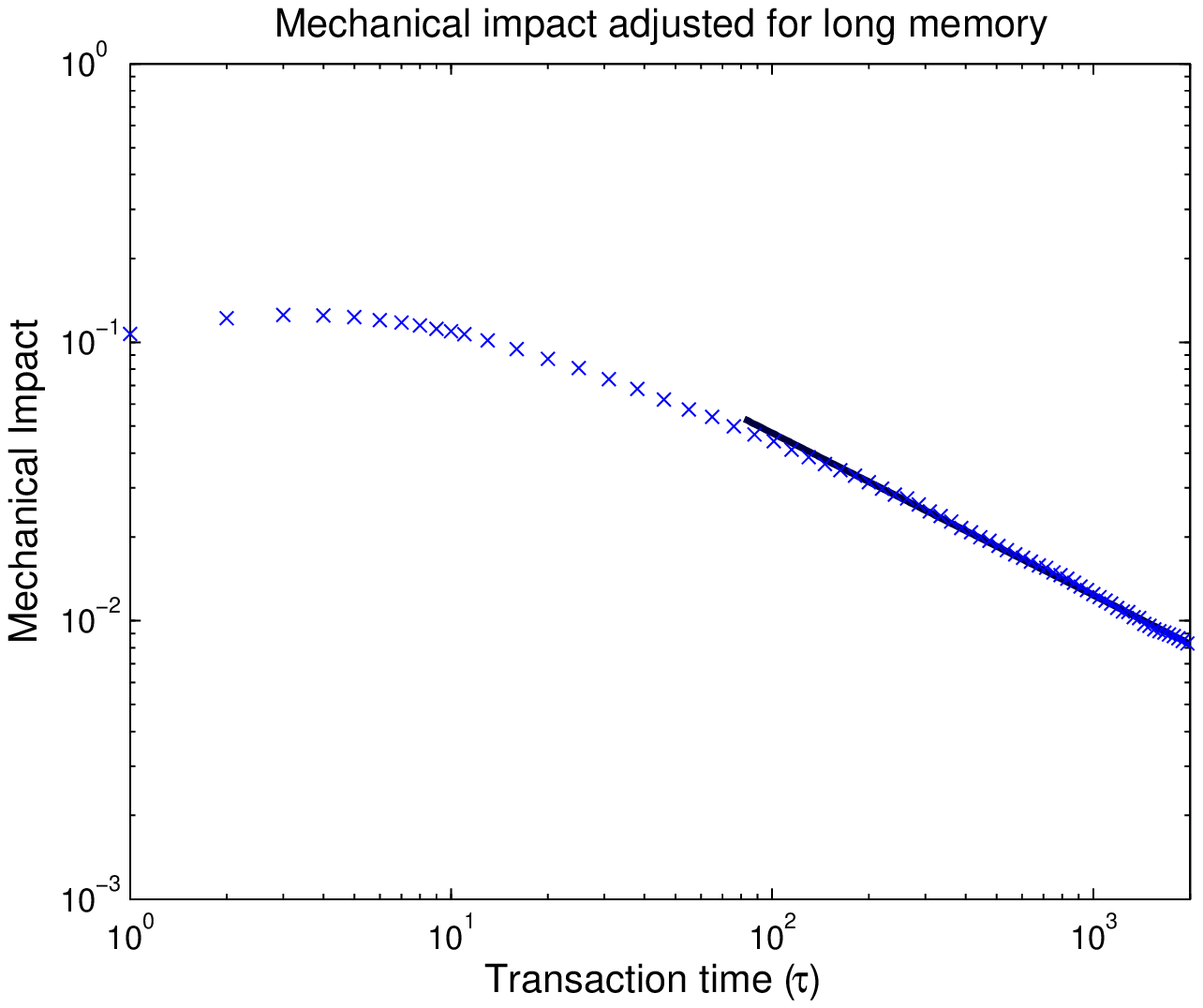}
\caption{(a) A comparison of the average mechanical, informational, and total impacts with and without adjustments for long-memory. In ascending order we show the average mechanical impact $R^M(\tau)$, the mechanical impact $C(\tau)$ adjusted for long-memory, the informational impact adjusted for long-memory $R^T(\tau) - C(\tau)$, the informational impact $R^T(\tau) - R^M(\tau)$, and the total impact $R^T(\tau)$, all as a function of time.  In (b) we show $C(\tau)$ over a longer time period in double logarithmic scale.  \label{cap:comparison-all-curves-zoomed}}
\end{figure}

When we compute the mechanical impact, because the perturbed transaction sequence and the reference sequence in Equation~\ref{mechanicalImpactDef} both have long memory, after subtracting them any effects due to long-memory disappear.  Nonetheless, it is clear that long-memory will amplify mechanical impact, for the same reasons it amplifies the bare impact.  Let the cumulative impact of an order at time $\tau$ be $C(\tau)$ and consider, for example, a buy order.  This generates a positive average impact at $\tau = 1$ of $R^M(1)$.  At $\tau = 2$ its average impact is reduced to $R^M(2)$, but we also need to take into account that the order at $\tau = 1$ is more likely to be a buy order than a sell order, so that there is an additional contribution $(P^+_1 - P^-_1) R^M(1)$.  Similarly at $\tau = 3$ the average cumulative impact $C(3) = R^M(3) + (P^+_1 - P^-_1) R^M(2) + (P^+_2 - P^-_2) R^M(1)$.  Noting that by definition $P^+_0 = 1$ and $P^-_0 = 0$, the general expression is
\begin{equation}
C(\tau) = \sum_{i=1}^{\tau} (P^{+}_{\tau-i} - P^{-}_{\tau-i})R^M(i)
\label{longMemCorrection}
\end{equation}
In Figure~\ref{cap:comparison-all-curves-zoomed}(a) we compare the raw mechanical impact, the cumulative mechanical impact $C(\tau)$ including amplification by long-memory, the informational impact, and the total impact.
Plotting the mechanical impact adjusted for long-memory on double logarithmic scale makes it clear that it is still decaying to zero, albeit slower than the raw mechanical impact. Fitting a power law to the long-time behavior of the form $K \tau^{-\eta}$, as shown in Figure~\ref{cap:comparison-all-curves-zoomed}(b) gives an exponent of about $\eta = 0.6$. 

In conclusion, while it is clear that the correlations associated with the long-memory of order signs amplify mechanical impact, they are not sufficiently strong to overcome the rapid decay of the mechanical impact to make it permanent.

\section{Correlation between mechanical and informational impacts\label{correlation}}

We now study the correlation between mechanical and informational impact.   Because mechanical impact is temporary, it is not obvious what feature of mechanical impact is likely to be most important.  Studying the correlation between mechanical and informational impact is motivated in part by the idea that prices are informative, i.e. as agents observe price changes induced by others, they update their own information.  Since mechanical impacts are temporary, it is not clear what others will respond to, particularly if the response is not instantaneous.  We somewhat arbitrarily use the integrated size, though one could easily argue that other properties of the mechanical impact might be more reasonable.  See the discussion in Section~\ref{causality}.

\begin{figure}[ptb]
\includegraphics[scale=0.6]{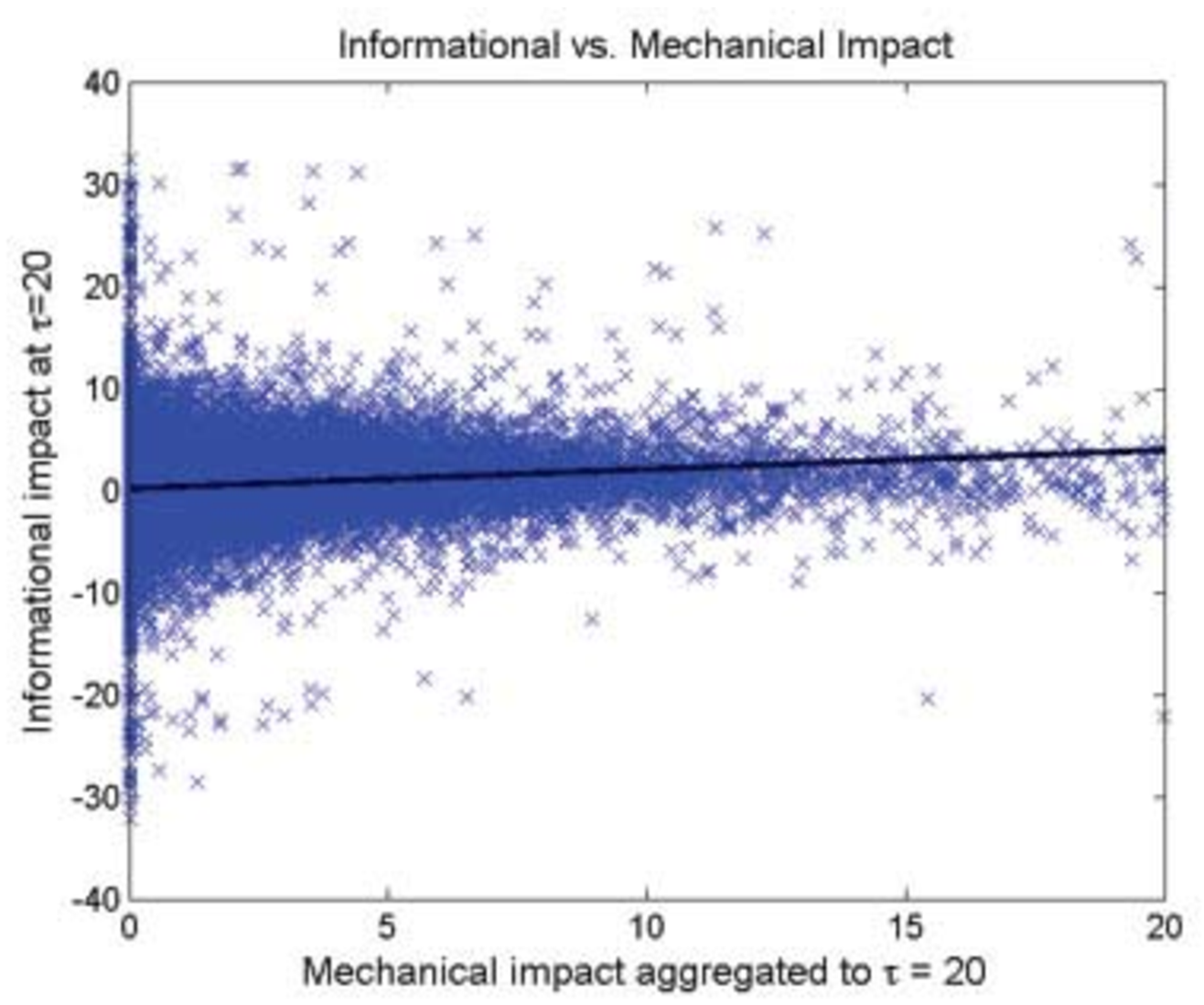}
\includegraphics[scale=0.6]{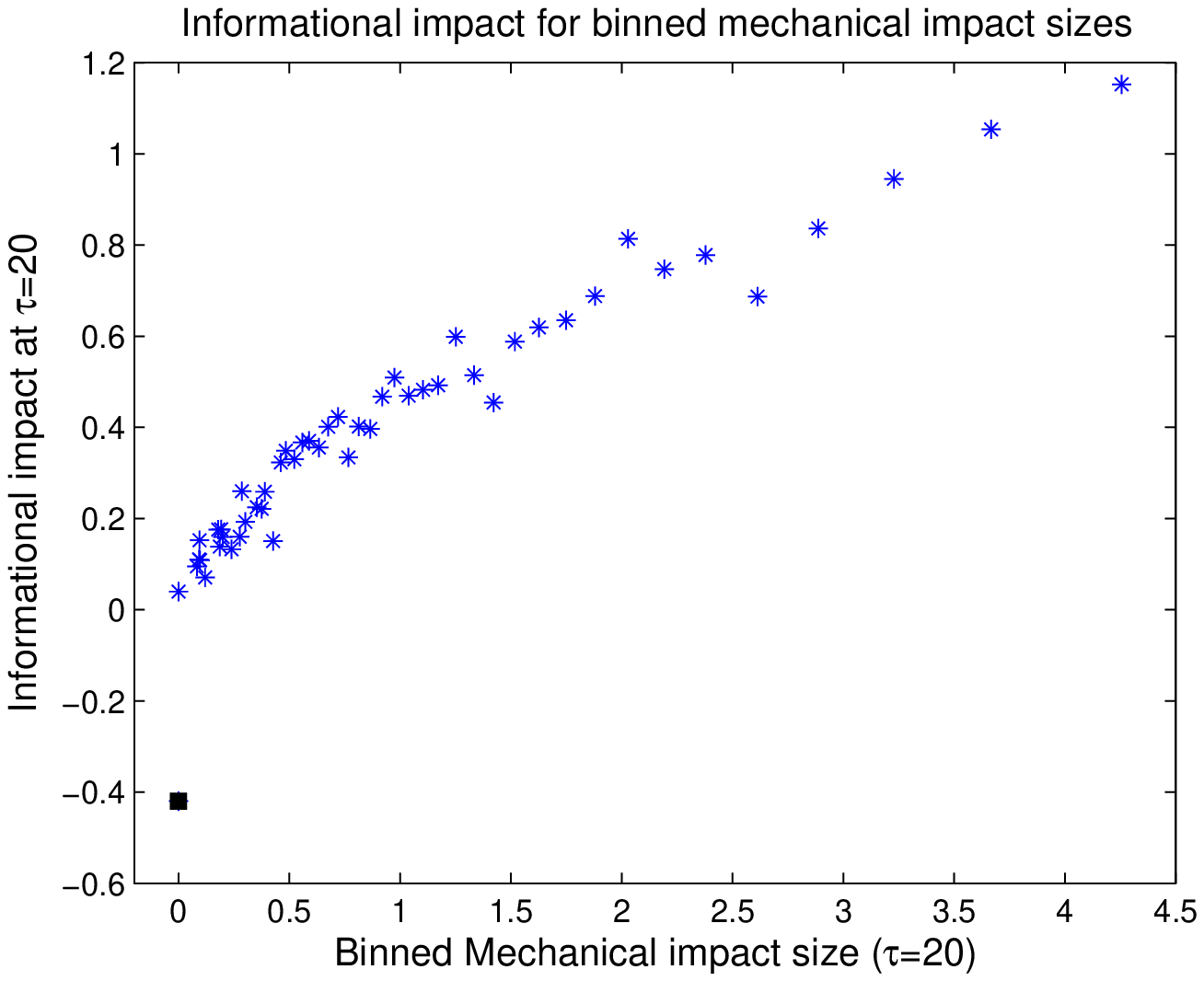}
\caption{The informational impact at $\tau = 20$ vs. mechanical impact integrated up to $\tau = 20$, $S_{20}^M (t)$.  In (a) the dots correspond to individual transactions, and the line is a regression using ordinary least squares.\label{cap:direct-vs-indirect-and-total}  In (b) the data is binned based on the integrated mechanical impacts, illustrating that even though the relationship is noisy, the correlation increases consistently with increasing size. Note that the black square corresponds to the case where the mechanical impact is strictly zero.}
\end{figure}

In Figure~\ref{cap:direct-vs-indirect-and-total} we plot the informational impact at $\tau = 20$ against the size of the mechanical impact integrated up to $\tau = 20$.    
The relationship for individual transactions is very noisy, but the positive association between mechanical and informational impact is quite clear.  The correlation is $\rho = 0.14$.  A linear regression of the form $\Delta p^I_{20} (t) = a S_{20} (t) + b$ yields a positive slope $a = 0.187 \pm 0.003$.  The error bar corresponds to a t-statistic of 55.  It is computed under the assumption of normally distributed IID data, and is certainly too optimistic, but it is nonetheless quite clear that the correlation is highly statistically significant.  Binning the data based on the integrated size of the mechanical impact shows that the relationship is essentially monotonically increasing -- large integrated mechanical impacts are associated with large informational impacts.  The positive association between informational and mechanical impacts is clearly highly statistically significant.

We find the surprising result that for cases where the mechanical impact is strictly zero, the informational impact has the opposite sign that one would normally expect, e.g. buy orders with strictly zero mechanical impact tend to have negative total price impacts. (Recall that when the mechanical impact is zero informational and total impact are the same). {\it Strictly zero} means that the mechanical impact is zero for all times in units of event-time rather than transaction time (the difference is situations where there is a non-zero impact in event time that dies out before the first transaction).  For AZN, for example, $13\%$ of effective market orders have strictly zero mechanical impact, in contrast to $33\%$ in transaction time.  To illustrate this in Figure~\ref{cap:direct-vs-indirect-and-total}(b) we have added the strictly zero case by placing two values over $0$ on the x-axis.  The first is consistent with the rest of the figure and corresponds to $S^M_{20} = 0$, while the square corresponds to cases where the mechanical impact is strictly zero.   The average informational impact in the latter case is  $\langle s_t \Delta p^I_{20} (t) \rangle_t = -0.4$, in contrast to the case where $S^M_{20} = 0$, which has roughly $\langle s_t \Delta p^I_{20} (t) \rangle_t = 0$.  Thus, as soon as there is any nonzero mechanical impact at all, the average informational impact jumps from a strongly negative value to a positive value.

\begin{figure}[ptb]
\includegraphics[scale=0.56]{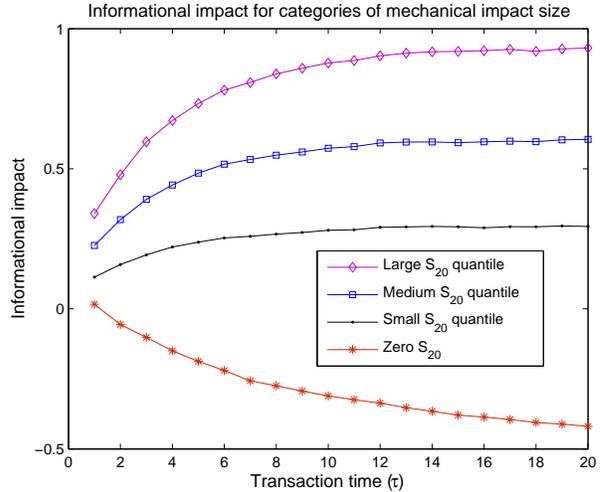}
\caption{Informational impact as a function of time for transactions with different levels of integrated mechanical impact. We show the average informational impact from $\tau = 1$ to $\tau = 20$ for three equal sized quantiles of the transactions with nonzero integrated mechanical impact $S_{20} (t)$, as well as for all transactions where the mechanical impact is strictly zero.   
\label{zeroVsNonZero}}
\end{figure}

The relationship between mechanical and informational impact remains consistent through time.  In Figure~\ref{zeroVsNonZero} we divide the mechanical impact into four different groups based on their mechanical impact and track the informational impact through time.  The first group consists of the cases whose impact is strictly zero and the others to the nonzero mechanical impacts, which are sorted based on $S_{20}^M$ into three groups with an equal number of events.
As before, the behavior of the informational impact when the mechanical impact is strictly zero is quite different than that when it is nonzero.  When the mechanical impact is strictly zero, the informational impact has a slightly positive value at $\tau = 1$, but then becomes increasingly negative as $\tau$ increases.  We hypothesize that this occurs because orders with no mechanical impact at all tend to be associated with price reversals.  For example, consider the case where we remove a buy market order which is too small to remove the best ask, in a situation in which the price immediately drops.  New sell orders will fill in below the best ask.  In this case it is very unlikely that removal of the buy market order will make any difference to the midprice.  In contrast, had the price continued to go up, the rearrangement in the level of sell limit orders triggered by the removal of the buy market order is much more likely to generate a perturbation in the price series.

\section{Summary\label{sec:Summary}}

We have introduced a precise definition of the mechanical impact of a transaction and demonstrated how it can be measured. The time behavior of the mechanical impact is highly variable. On average, however, for market orders it is strongest immediately after the order is placed, and then decays to zero over time. It is initially the dominant component of the total price impact, but the relative fraction decreases as the mechanical impact shrinks and the informational impact grows to approach its asymptotic value.  Initially the mechanical impact is about half as large as the asymptotic informational impact.

\subsection{Causality\label{causality}}

We have defined the informational impact as what is left over of the total impact once the purely mechanical impact is removed.  The justification for this is that the remainder depends on correlations between events.  I.e., unless there is something correlating the information in the reference event $\omega_t$ to the subsequent sequence of events $\Omega_{t+1}$, there will be no informational impact.  This leaves open the question of causality. One hypothesis is that event $\omega_t$ causes changes in the subsequent events $\Omega_{t+1}$. The alternative hypothesis is that there is a common cause for $\omega_t$ and any aspects of $\Omega_{t+1}$ that are correlated with its presence. Based on the information available here we cannot distinguish these two hypotheses.

The idea of a common cause is easy to understand. Suppose, for example, that an external event causes a group of investors to decide to buy, and they submit a series of buy orders without paying any attention to each other. This will cause a rise in price associated with each buy order, due both to the mechanical impact of each buy order, and the mechanical impact of all the buy orders that are correlated with it. The correlated part of the price rise will be measured as an informational impact (reflecting the external information).

The causal hypothesis is more interesting, and is connected to the role of trading and its associated price changes in transmitting information. In a world where individual agents have private information, trading provides a mechanism for disseminating that information. If an agent receives new private information, this may cause him or her to trade. Trading affects the price, which is visible to everyone.  An intelligent agent with different information will observe the change in price, and will infer that his or her valuation must be wrong. As a result, each agent will arrive at a valuation that is based partly on private information and partly on price and other public information.  Prices are thus a mechanism for making private information public.  This idea is well-accepted in economic theory \cite{Grossman80,Grossman89}.  

Insofar as the causal hypothesis is correct, the temporary nature of mechanical impact suggests that private information is made public through a highly dynamic process. As each trade happens, it causes a mechanical impact, which is a signal visible to all. Before it decays away, it can cause other trades of the same sign to occur, or it can cause cancellations to occur, or (probably most important) it can cause changes in the limit prices of subsequent orders. Each of these events has its own mechanical impact, creating a cascade of impacts giving rise to a permanent change. The avalanche-like nature of this process suggests possible analogies to self-organized criticality.

At this stage we are unable to say to what extent common cause or causality are at play. It seems likely that both are acting. We do know that these series display long-memory, and as discussed in Section~\ref{longMemory}, evidence suggests that this is essentially an exogenous phenomenon, which from this point of view is a common cause. 

\subsection{Generalizations\label{generalizations}}

The mechanical impact can be defined in a way that is more general than what we have presented here.  We have defined mechanical impact by fully removing an event, but there are many other possible modifications of the order book that are worth considering.  Equation~\ref{mechanicalImpactDef} can be generalized to read
\begin{equation}
\Delta p^M_\tau (t) = \Pi(b_t, \Omega_{t+1}^{t + \tau}) - \Pi(\tilde{b}_t, \Omega_{t+1}^{t + \tau})
\end{equation}
where $b_t$ is the real order book at time $t$ and $\tilde{b}_t$ is any modification of it that may yield a useful interpretation.  For example, $\tilde{b}_t$ might be the true order book with an additional order added, or it could be an infinitesimal modification of a given order, e.g. with the size of the most recent order $\omega_t$ slightly enhanced or diminished.  This potentially allows one to make a quasi-continuous analysis, effectively estimating the derivative of the mechanical changes in prices with respect to changes in supply or demand.

\subsection{Future work}

We have left many questions unanswered in this work. In addition to the generalization discussed above, there are many topics that remain to be investigated. For example, what is the dependence of mechanical impact on the size of orders?  Previous results have shown that the total impact is remarkably independent of order size \cite{Farmer04b,Weber04}; is this also true for mechanical impact?  Is the mechanical impact symmetric under increases or decreases in the size of $\omega_t$?  What does the price impact of cancellation and limit orders look like?  We hope to investigate these and other questions in the future.

\acknowledgments We would like to thank Barclays Bank and the Capital Markets CRC for supporting this research, J.-P. Bouchaud, Marc Potters, Austin Gerig, and Adlar Kim for useful comments, and the London Stock Exchange for supplying data.

\end{document}